\documentclass{elsart}
\usepackage{epsfig}
\usepackage{natbib}
\begin{document}

\begin{frontmatter}



\title{Study of a detector array for Upward Tau Air-Showers}


%\author{} 
%
%\address{}

\author[Rome]{M.~Iori\thanksref{1}},
\author[Rome]{A.~Sergi} and
\author[Rome]{D.~Fargion},
\author[Rock]{M. Gallinaro},
\author[kafkas]{M. Kaya}
\address[Rome]{University of Rome ``La Sapienza'' and INFN, Rome, Italy}
\address[Rock]{Rockfeller University, New York, USA}
\address[kafkas] {University of Kafkas, Kars, Turkey}
\thanks[1]{maurizio.iori@roma1.infn.it}

\begin{abstract}
The cosmic ray spectrum extends to energies above 10$^{20}$~eV. In
direct production or acceleration models, as well as by photo-pion
interaction  high energy  cosmic ray flux must contain neutrinos
and photons. The latter are absorbed by cosmic radiations while
neutrinos are not. The need of a Neutrino Astronomy is compelling.
In this paper a study of a detector array designed to measure horizontal $\tau$
air-showers emerging from the ground, produced by $\nu_{\tau}$
interactions with the Earth's crust, is presented. Each array unit is
composed of a pair of scintillator tiles mounted on a frame with a
front field of view of about 0.1$~sr$, optimized to distinguish
between up-going and down-going crossing particles by their
time of flight. The detector array sensitivity, the size of the
array and the $\tau$ shower identification are discussed. Because
of the almost complete mixing of $\nu _{\mu} \to \nu _{\tau}$
the ultrahigh energy neutrino tau and its minimal consequent
tau-airshower rate is estimated; 
assuming that the neutrino energy spectrum follows a
Fermi-like power law
$E^{-2}$, the sensitivity with 3 years
of observation is estimated to be about 60 eV cm$^{-2}$s$^{-1}$sr$^{-1}$
in the energy range $10^{17-20}$ eV.
This value would provide competitive upper limit with present and future
experiments. We found also that, in the same time, this system can observe about one GZK neutrino
event per km$^2$.
\end{abstract}
\end{frontmatter}

%\input{psfig}

% main text
%\section{}
%\label{}
\section{Introduction}

In the rising era of UHECR  astronomy, the understanding of
astrophysical phenomena can shed light on the structure and
evolution of the Universe. In particular, the observation of
ultra-high energy cosmic rays (UHECRs) has raised new questions
about their source and their composition. The old Ginzburg
scheme~\cite{Ginzburg}, according to which cosmic rays come from
galactic sources such as supernova remnants or pulsars, has
received new attention as they can be possible neutrino
sources~\cite{R1,R2}. However Supernova may provide up to PeVs
energy accelerations. Cosmic rays are mainly protons and atomic
nuclei that are observed to strike the Earth with exceedingly
large energies, but lower and lower fluxes. Their energy spectrum
extends to energies above $10^{20}$eV and is mostly well described
by a power law with two kinks, a ``knee'', where the slope
steepens at energies of about $3\times10^{15}$eV, and an ``ankle''
at energies of $3\times10^{18}$eV, where the spectrum flattens. It
is generally accepted that most cosmic rays below the knee may
come from inside our galaxy while highest energy ones, above the
ankle, from extra-galactic sources. However, the origin of
UHECR's, i.e. those above the knee, remains a puzzling mystery.
Nobody knows with security where they are produced, and no
guaranteed source has  been detected yet. The idea that
high-energy cosmic rays originating in our galaxy and those
originating from extra-galactic sources could come from the same
kind of source has been suggested~\cite{cr_origin,arons}. The
origin of UHECRs above the ankle could be due to particles
generated in the cores of Active Galactic Nuclei (AGN), which
could then produce neutrinos through interactions of accelerated
protons and ions with an accretion disk target surrounding a
central black hole \cite{R3}. In analogy UHECR has been suspected
to be produced by GRBs and SGRs  sources. However, the UHECR flux
is expected to be suppressed after interacting with the cosmic
microwave background radiation when cosmic rays have energies
above $4\div 5~10^{19}$~eV. Such a rate drop is generally referred
to as the ``GZK cutoff''~\cite{GZK}, following the
Greisen, Zatsepin and Kuzmin proposal. If the GZK cutoff is not
confirmed by experiments, new physics must be invoked to
understand the origin of these extreme events, and the neutrinos
could play a crucial role both as a source \cite{farg0}
as well as a probe.

Indeed neutrinos are often associated with high energy cosmic ray
fluxes and, since they interact only weakly, they carry
directionality as a characteristic signature of their source.
Therefore, high energy neutrino observations are valuable probes
of new astronomy. Information is encoded in the energy spectrum,
arrival direction, and flavor content of the cosmic neutrinos. The
measurement of the neutrino-flavor mixing rate is also important
to determine the origin of these particles. These UHE particles
may come from very rare exotic sources due to dark matter decay
\cite{sigl} and annihilations~\cite{DFK,neutral}.

Most experiments can detect neutrinos by looking for the most
penetrating up-going lepton tracks (the muons) produced by charged
current interaction of parental neutrinos with energies of
$10^{11}-10^{16}$~eV. Neutrino atmospheric secondaries from common
cosmic ray flux  in this energy range are investigated using several
techniques~\cite{amanda,ice,sudan}.

However a totally novel technique has been proposed based on the
peculiar tau decay in flight: the discovery of UHE tau neutrinos
at Horizons or Upgoing (Hortaus and Uptaus or Earth-Skimming
Neutrinos)\cite{FAC,farg0,Fargion,farg2,Feng} whose
powerfull air-shower tests the PeV-EeV tau neutrino astronomy. The
up-going signature is noise free respect to downward ones.

Indeed the on-going experiments,
which are designed to detect particle showers with energies above
$10^{19}$~eV and are sensitive to neutrinos coming from an almost
horizontal direction. For example, the Southern site of the Pierre
Auger Observatory is able to detect down-going air-showers through
photo-luminescent light at azimuthal angles of $\theta <
60^\circ$, and has also a large acceptance to the showers produced
by Earth-skimming neutrinos with energies above $10^{19}$eV
\cite{Auger1,Auger2}.

None of the experiments devoted to the exploration of
neutrino physics or UHECRs, such as Amanda/Icecube~\cite{amanda,ice}, Antares, Magic or NuTel,
are specifically designed to measure the $\nu_\tau$
flux in the energy region above $10^{18}$eV with a large duty cycle~\cite{zas}.
The proposed detector array described here is designed with improved sensitivity to measure or
set an upper limit on the flux of Earth-skimming tau neutrinos.
This measurement will introduce new
constraints on the flux of UHECRs,
and distinguish between the AGN and the GZK production mechanisms.

The paper is organized as following.
In Section~\ref{sec:idea}, the results of simulation studies of atmospheric hadron and inclined $\tau$ showers are presented;
configuration of the detector array, detector acceptance and predicted event rates are discussed in Sections~\ref{sec:detarray}-\ref{sec:accdet}.
In Sections~\ref{sec:tausel}-\ref{sec:taurec}, the selection and the reconstruction of the tau showers are studied;
Section~\ref{sec:trigger} contains a discussion of the trigger.
In Section~\ref{sec:prototype} the results obtained with a 2-detector
array prototype installed at the High Altitude Jungfraujoch
Laboratory (3600~m~a.s.l., Switzerland) are presented.

\section{Showers produced by UHE hadrons and $\nu _{\tau}$'s }
\label{sec:idea}

\begin{figure}[!htbp]
\begin{center}
\includegraphics[angle=270,width=0.7\textwidth]{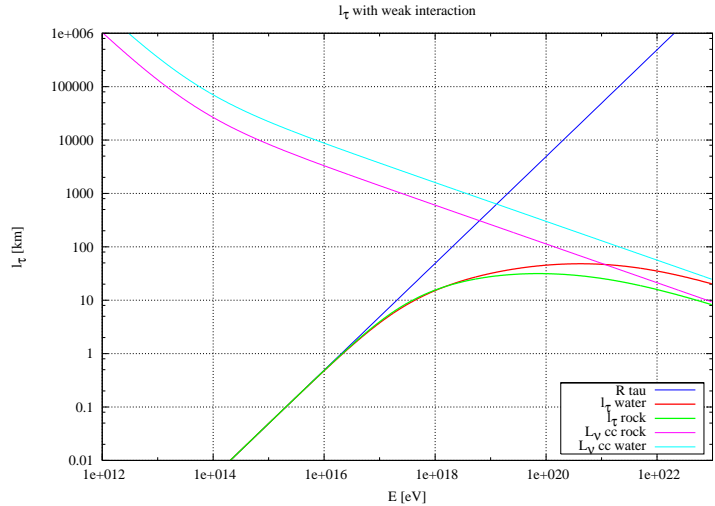}
\caption{\label{fig:taurange} The neutrino interaction length as a function
of the energy for in water and rock and tau range in air, water and rock \cite{farg2}. }
\end{center}
\end{figure}

The ability of detecting tau neutrinos increases significantly for
neutrinos that enter the atmosphere at large zenith angles, at
horizontal or almost below the horizontal direction, traversing a
small section of the Earth's crust. In fact, at energies above
$10^{18}$~eV the entire Earth becomes opaque to neutrinos and only
horizontal or Earth-skimming high-energy neutrinos can be
detected. Neutrinos coming from a vertical direction have a
negligible probability to interact with the atmosphere (a ten
meter water equivalent screen) and remain mostly invisible to
detectors above the ground. When Earth-skimming neutrinos
interact, they may generate tau whose interaction lenghts may
exceed  range in rock of 5-10~km at energies above 
$10^{17}$~eV (Fig.~\ref{fig:taurange}) \cite{farg2}.

The ratio of the interaction length of the tau traversing the rock divided by the interaction length of the neutrino,
$\lambda_{\tau -rock}$/$\lambda_{\nu}$, is about 1\% in the energy interval of $10^{18-19}$eV.

The main source of background to the neutrino flux comes from atmospheric protons and muons, and
the most difficult experimental challenge comes from separating this background from inclined tau showers.
The aim of the study presented in this section is to investigate the shower properties
using a Monte Carlo simulation.
In fact, the topology of the shower can be used to distinguish UHE neutrinos from the ordinary atmospheric
hadron interactions~\cite{Fargion,Feng,Beacom,Cao}.
%UHE neutrinos can be detected and distinguished from the ordinary atmospheric
%hadron interactions by the topology of the shower~\cite{Fargion,Feng,Beacom,Cao}.

Ordinary hadrons interact when entering the atmosphere and initiate down-going air-showers.
At large zenith angles ($\theta\simeq 90^\circ$) they
interact at a distance of about 400~km from the ground, with a maximum development of the
shower at approximately 100~km after the interaction occurs.
Because of interactions with the atmosphere, at the ground level, most of the electromagnetic component of the shower
disappears and only more energetic muons survive.
The density of muons coming from protons with an initial energy of $5\times10^{18}$eV
is estimated using the Aires simulation~\cite{Aires} to be of the
order of $10^{-2}$ muons/m$^2$.
On the other hand, UHE neutrinos interact weakly with protons and nuclei through charged or neutral currents
producing leptons or neutrinos with energies of $10^{17-20}$eV.
When a $\nu _{\tau}$ interacts through charged current, high energy $\tau$ leptons are produced with long
decay lengths in the range of 5-50 km
 which then originate air showers with a small transverse size (a few kms), if occurring near the observer.
The arrival direction of the $\nu_\tau$ can also be inferred from the shape of the shower itself.

Simulation studies suggest that the time structure and the muon density significantly differs, at ground level
in specific geometrical conditions, in $\tau$ and  hadron showers (see section~\ref{sec:tausel}).
In this study only the tau hadronic decay modes
(for instance, $\tau \to \pi \pi \pi \nu_{\tau}$) have been considered, which are about 64\% of
all tau decay modes.
The shower initiated by a tau lepton decaying into $\pi \pi \pi \nu_{\tau}$
has a density of about 10$^{3}$ times larger than a horizontal atmospheric
proton shower of the same energy and zenith angle due to the long path in the
atmosphere.

\section{Design of the detector array}
\label{sec:detarray}

\begin{figure}[!htbp]
\begin{center}\framebox{
\includegraphics[width=8cm]{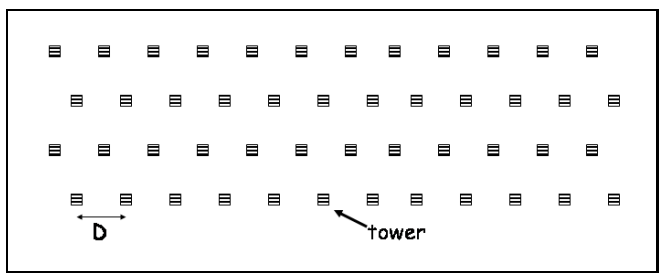}}
\caption{\label{fig:DetArraysketch}
Schematic aerial top view of a partial detector array. Four rows of twelve towers each are spaced
by a distance $D$ (not to scale).}
\end{center}
\end{figure}

In order to detect muons produced by inclined air showers, several detectors pointing
at the horizon and arranged in a large surface array must be assembled.
The elementary modules, called {\it towers}~\cite{mine},
can be arranged in a grid with a spacing of distance $D$ (Fig.~\ref{fig:DetArraysketch}).
In a grid with detector spacing $D$=100~m, for example,
the time correlation of the signals from the individual towers of about 300~ns
allows the identification of the arrival direction of the shower.

\begin{figure}[!htbp]
\begin{center}\framebox{
\includegraphics[width=12cm]{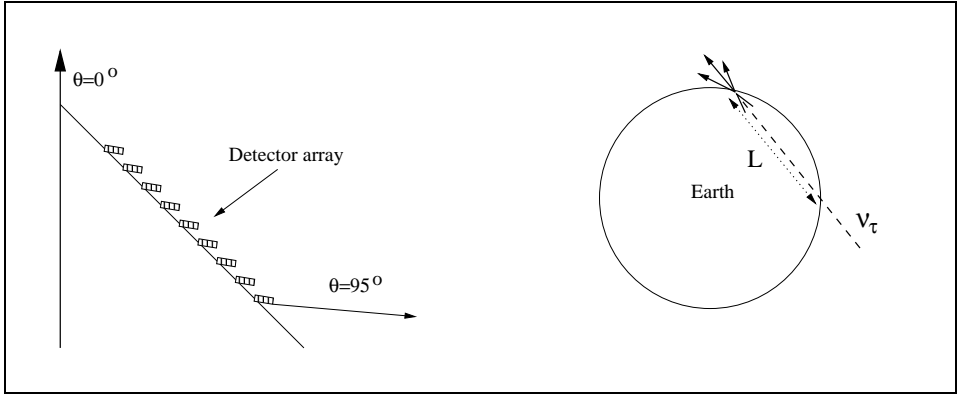}}
\caption{\label{fig:detectorarray}
Schematic drawing (not to scale) of the detector array located on the slope of a mountain along a zenith angle of $95^\circ$ (left),
and pointing at the incoming inclined showers from Earth's skimming neutrinos (right),
after traversing a distance $L$ through the Earth's crust.}
\end{center}
\end{figure}

Due to the fact that the Earth-skimming neutrinos are coming from inclined showers,
detector arrays pointing at the horizontal direction are more efficient than those pointing at the sky at small zenith angles.
Therefore, the location of the detector array, pointing downward on an inclined plane or a mountain slope
at large elevation to cover a large solid angle,
can substantially increase the detector's acceptance and improve the rejection of background events
(Fig.~\ref{fig:detectorarray}).
Ground detector arrays have small trigger efficiency for inclined showers
close to the $90^\circ$ zenith angle.
They can mostly detect horizontal $\tau$ showers generated in the atmosphere.

In order to measure upward moving showers coming from $\nu _{\tau}$ interactions with the Earth's crust,
the detector
must be able to distinguish between up-going and down-going particles,
and towers can be individually used for this purpose.
Each tower (Fig.~\ref{fig:tower}) consists of two parallel scintillator plates placed at a distance of 160~cm
and is instrumented with a precision timing measurement device that allows to measure
the time of impact of a particle passing through the scintillator tiles.
The particle direction can be determined and up-going particles can be selected.
The charge deposited can also be measured.
Results of a tower prototype designed to measure the horizontal flux have been previously reported~\cite{mine}.
Further studies to determine the horizontal flux from atmospheric showers
have been performed at the Jungfraujoch Laboratory, which is located at 3600~m a.s.l.,
and are reported in section~\ref{sec:prototype}.

The shower shape at ground level has been studied to determine
the needed granularity of the detector array, i.e. the distance between towers.
The muon density of a $5\times 10^{18}$eV $\tau$ shower emerging from the ground
is about 10 particles/m$^2$, in average, and a minimum of 2 particles/m$^2$
at the border of the shower;
therefore, using a tile size of $20\times20$~cm$^2$ and assuming a constant
shower density,
each tower has a probability of 40\% and 8\% to detect a particle in
the core and at the border of the shower, respectively.

Due to the relatively short longitudinal development of the shower~\cite{Gaisser},
the distance between the ground, from where the shower is emerging, and the detector array should be kept short
and preferably about 10~km. % (Fig.~\ref{fig:effDL}).
At increasing ground-to-detector distances the contamination from background events coming from above the horizon also increases.
No $\nu_\tau$ regeneration from $\tau$ decays has been taken into
account~\cite{Kolb}.
The study of the shape of a $\tau$ shower at ground level also indicates that two detector
arrays located on two nearby hills or slopes of a mountain may improve
the acceptance, i.e. using two rectangular arrays separated by a gap.

\section{Detector acceptance and predicted event rate}
\label{sec:accdet}

\begin{figure}[!htbp]
\begin{center}
\includegraphics[width=6.cm,totalheight=7.cm]{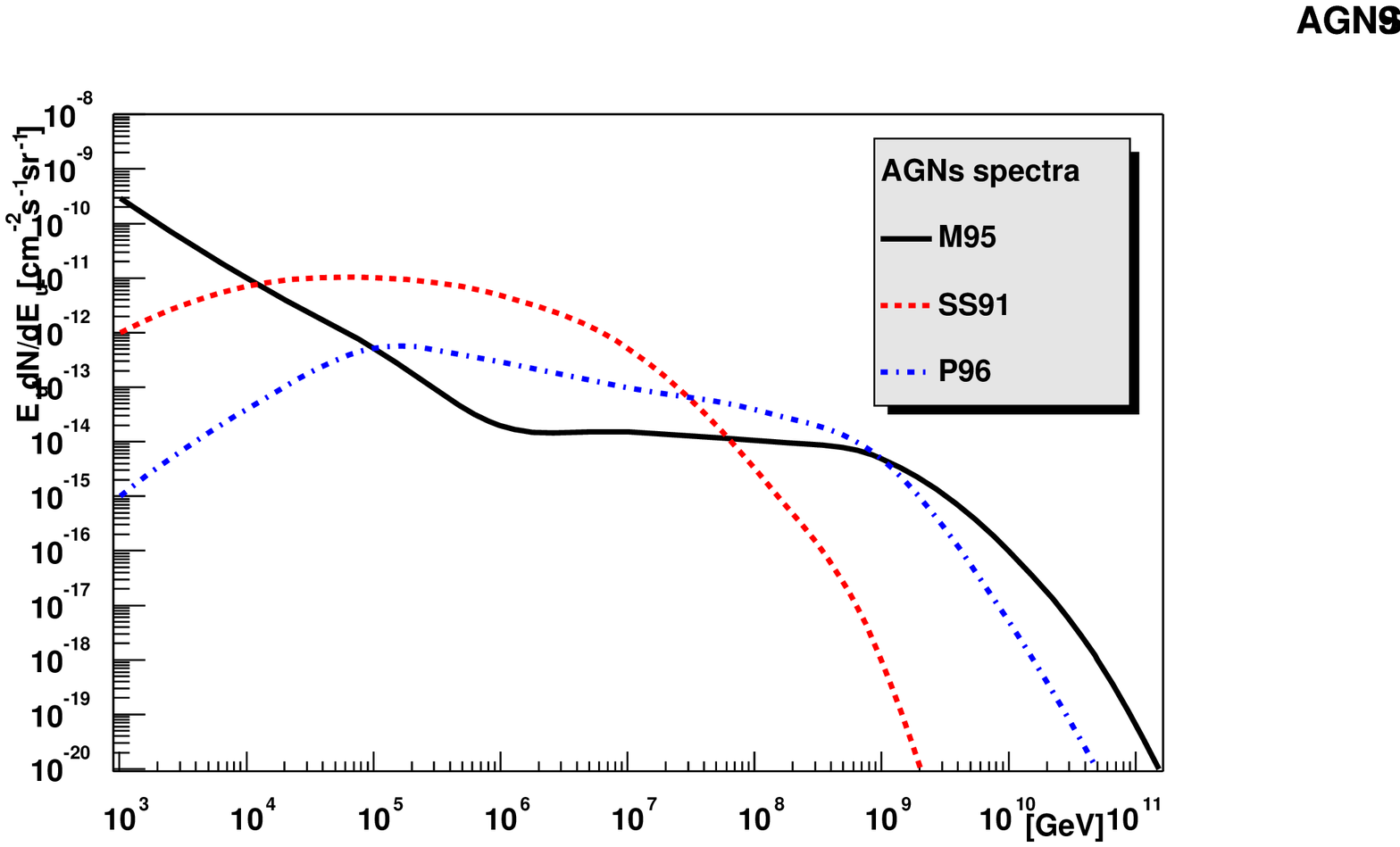}
\includegraphics[width=6.cm,totalheight=7.cm]{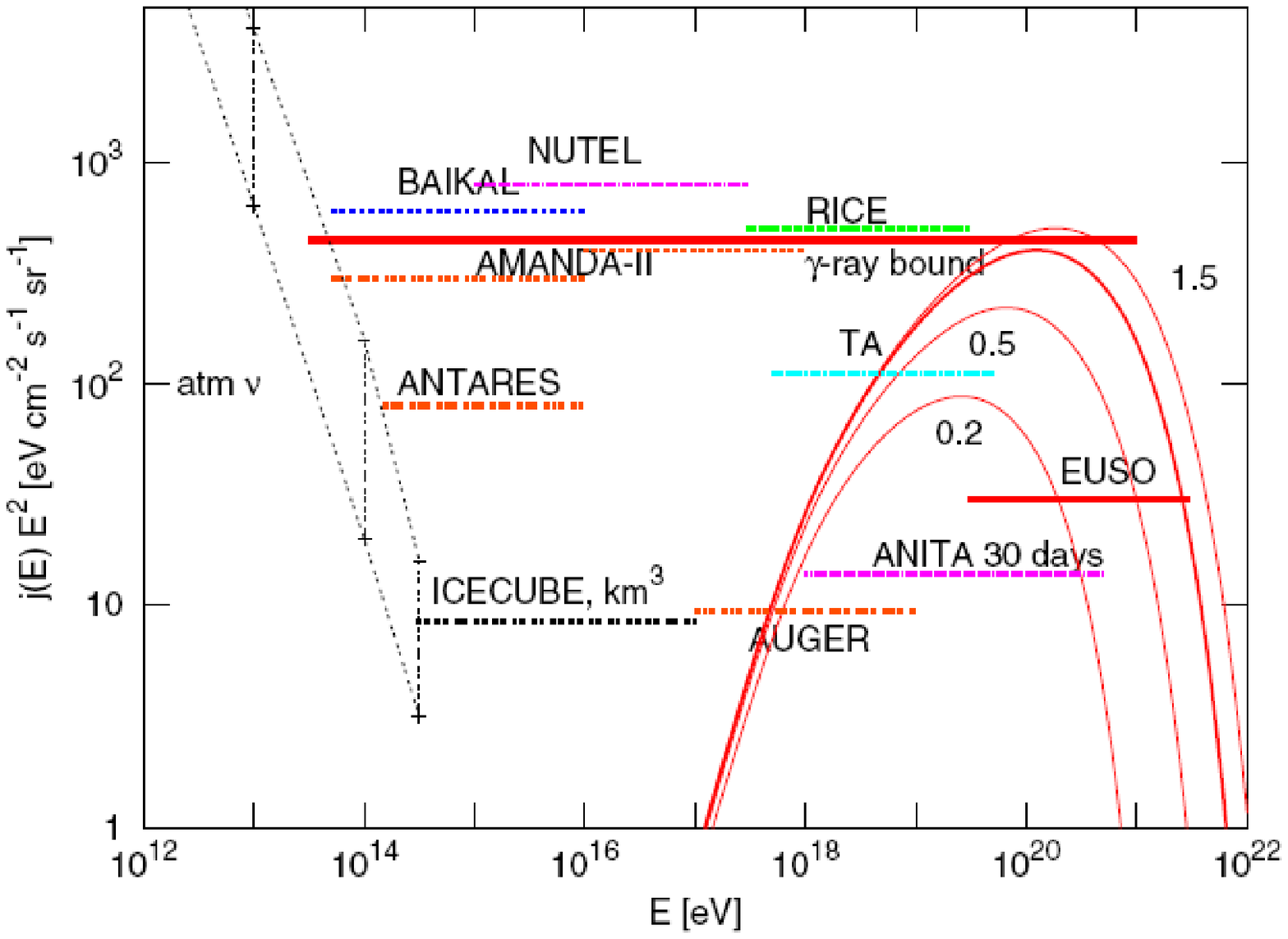}
\caption{\label{fig:AGN}
Energy spectra of the predicted neutrino flux from the AGN model: M95\cite{M95}, P96\cite{P96} and SS91\cite{SS91} (left);
cosmological neutrino flux compared with expected
sensitivity to tau neutrino flux for the planned projects (right).
From Ref.\cite{Dmitry,AGN}.
}
\end{center}
\end{figure}

The flux of $10^{18}$eV $\nu_{\tau}$ is estimated to be of the order
of 10~km$^{-2}$yr$^{-1}$sr$^{-1}$ in the GZK model
and about $10^2$ larger in the AGN production mechanism
(Fig.~\ref{fig:AGN}, left).
According to these expectations, in order to detect at least 1 event per year,
the detector acceptance, defined as \emph{detector sensitivity}, must be of the order of 1-10~km$^{2}$sr.

\begin{figure}[!htbp]
\begin{center}
\includegraphics[width=10cm,totalheight=8cm]{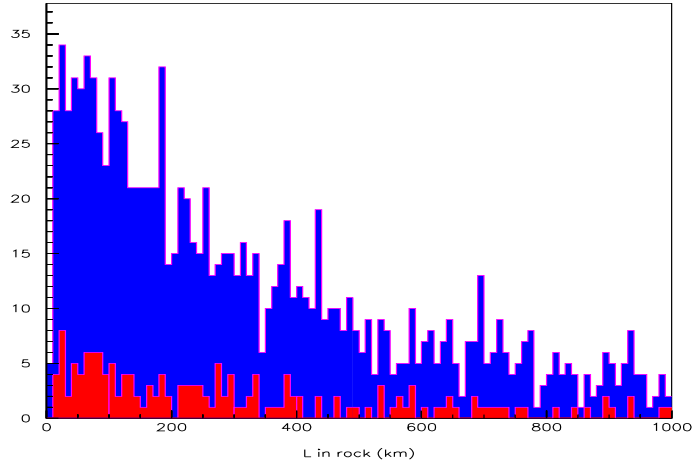}
\caption{\label{fig:chord} Neutrino interactions along the Earth's crust
where a tau reaches the ground (shaded blue). The red shaded area is the
fraction of tau emerging the ground and decay with the maximum of its shower
 on the detector plane distant 10 km from the emerging point. For $\nu_\tau$ flux
is considered a power law $E^{-2}$ in the energy range $10^{17}-10^{20}eV$.
}
\end{center}
\end{figure}

\begin{figure}[!htbp]
\begin{center}
\includegraphics[width=10.cm,totalheight=8.cm]{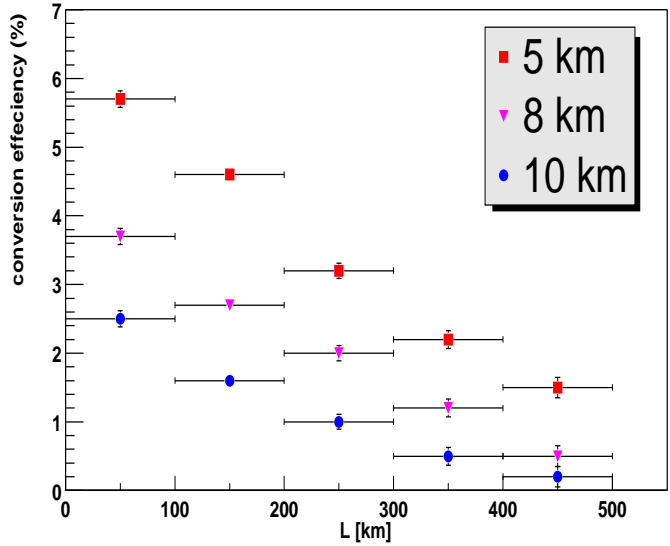}
\caption{\label{fig:accdet}
Conversion efficiency for different ground-to-detector distances
(5,8,10~ km) as function of distance $L$, the combined path of $\nu_\tau$ and $\tau$
through the Earth's crust. For $\nu_\tau$ flux
is considered a power law $E^{-2}$ in the energy range $10^{17}-10^{20}eV$.}
\end{center}
\end{figure}

The probability that a $10^{18}$eV $\nu_\tau$ produces a 10$^{17}$eV tau
emerging from the ground depends of the distance $L$ traversed through the Earth's crust
(Fig.~\ref{fig:detectorarray}) and on the tau energy loss in the rock.
This probability is calculated as the convolution of
the probability that the $\nu_{\tau}$ interacts before a distance $L$
with the probability that the $\tau$ reaches a distance $L$.
Figure~\ref{fig:chord} shows, considering $\nu_\tau$ flux with $E^{-2}$ power law
in the energy range $10^{17}-10^{20}$eV, the number of taus
reaching the ground,
which decreases as the distance $L$ increases.
For L=200-300~km, which corresponds to a zenith angle of 92.5$^\circ$,
the conversion efficiency varies from 1.5\% to 3\% as function of L
along the length of the detector array between 10~km and 5~km, respectively (Fig.~\ref{fig:accdet}).
The conversion efficiency is defined as the probability of getting a tau shower
at ground level having a shower maximum close to the detector plane.

\begin{figure}[htbp]
\begin{center}
\includegraphics[height=4cm]{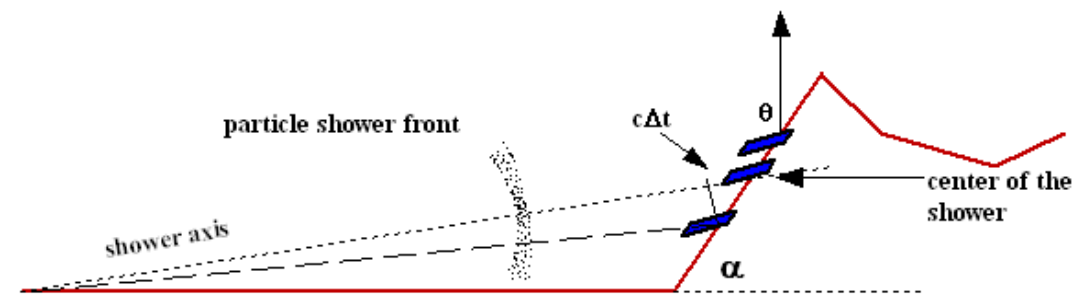}
\caption{\label{fig:mont} Schematic drawing of the particle shower front impinging on the detector array.
}
\end{center}
\end{figure}

The predicted event rate is calculated for
a detector array located on a surface inclined at an angle $\alpha$ with respect to the ground (Fig.~\ref{fig:mont}).
The solid angle can be improved if the detector array is composed by $units$,
each formed by two towers, shown in Fig.\ref{fig:tower} and positioned nearby (i.e.~60~cm) and parallel to each other.
In this case,
if each unit has opening angles $\Delta \phi=40^\circ$
and $\Delta \theta=15^\circ$
and $S_{eff}$ is
the effective area defined like the projection of the
detector plane to the transverse shower section,
we obtain an acceptance of about
$S_{eff}$0.15 $km ^{2} sr$.
Assuming that the energy spectrum of cosmological neutrinos follows
the power law
$d\Phi/dE_\nu=10^{-6}E_\nu^{-2}$GeV$^{-1}$cm$^{-2}$s$^{-1}$sr$^{-1}$,
an integrated flux of $3\cdot 10^3$km$^{-2}$yr$^{-1}$sr$^{-1}$
is expected in the energy range $10^{17}-10^{20}$eV,
which corresponds to approximately 5 events per yr km$^{2}$.
Assuming that the neutrino energy spectrum follows a Fermi-like power law
$E^{-2}$, the sensitivity with 3 years
of observation is estimated to be about 60 eV cm$^{-2}$s$^{-1}$sr$^{-1}$
in the energy range $10^{17-20}$ eV.
This value would provide competitive upper limit with present and future
experiments. We found also that, in the same time, this system can observe about one GZK neutrino
event per km$^2$.
%If instead, the neutrino energy spectrum falls as
%$d\Phi/dE_\nu=10^{-7}E_\nu^{-2}$GeV$^{-1}$cm$^{-2}$s$^{-1}$sr$^{-1}$
%only an upper limit on neutrino production can be set.
%A flux of about $10$km$^{-2}$yr$^{-1}$sr$^{-1}$, which is expected in the GZK framework,
%would yield no candidate events in similar experimental conditions.
It is therefore possible to exclude or confirm several models (AGN-M95\cite{AGN-95J}, MPR\cite{MPR})
and determine the energy spectrum of UHE neutrinos.

\section{Identification of a tau shower}
\label{sec:tausel}

\begin{figure}[!htbp]
\begin{center}
\includegraphics[width=8cm]{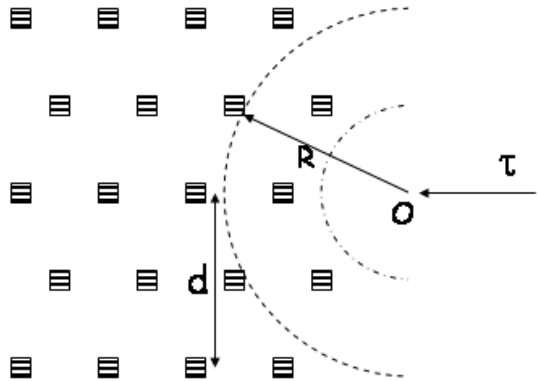}
\caption{
\label{fig:wavefront_sketch}
Schematic drawing of the time propagation through the detector array of a shower originating in the origin $O$
which exhibits a curvature radius $R$ at detector level;
$d$ is the distance from the axis of the shower direction.}
\end{center}
\end{figure}

\begin{figure}[!htbp]
\begin{center}
\includegraphics[width=6.cm,totalheight=7.cm]{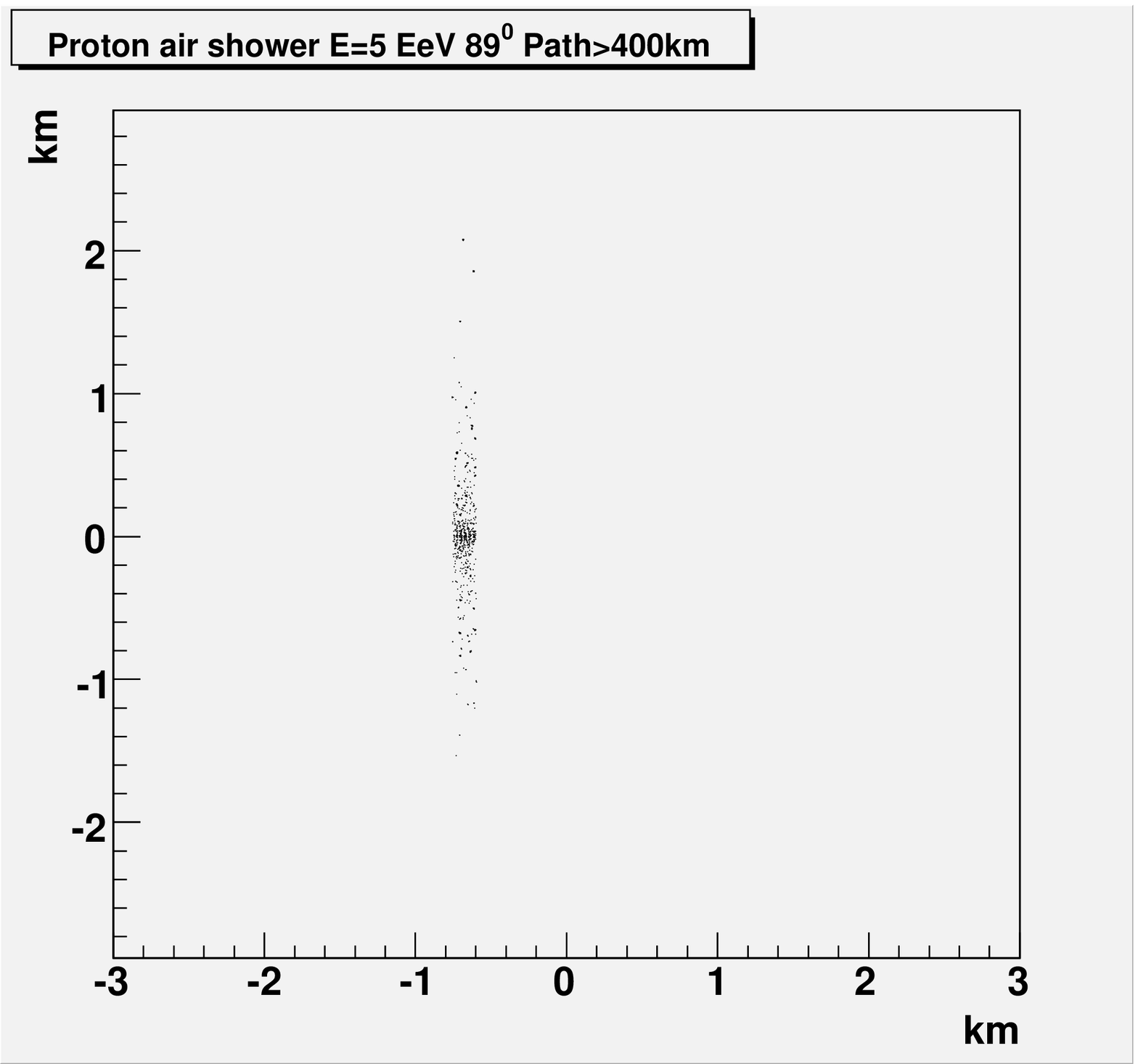}
\includegraphics[width=6.cm,totalheight=7.cm]{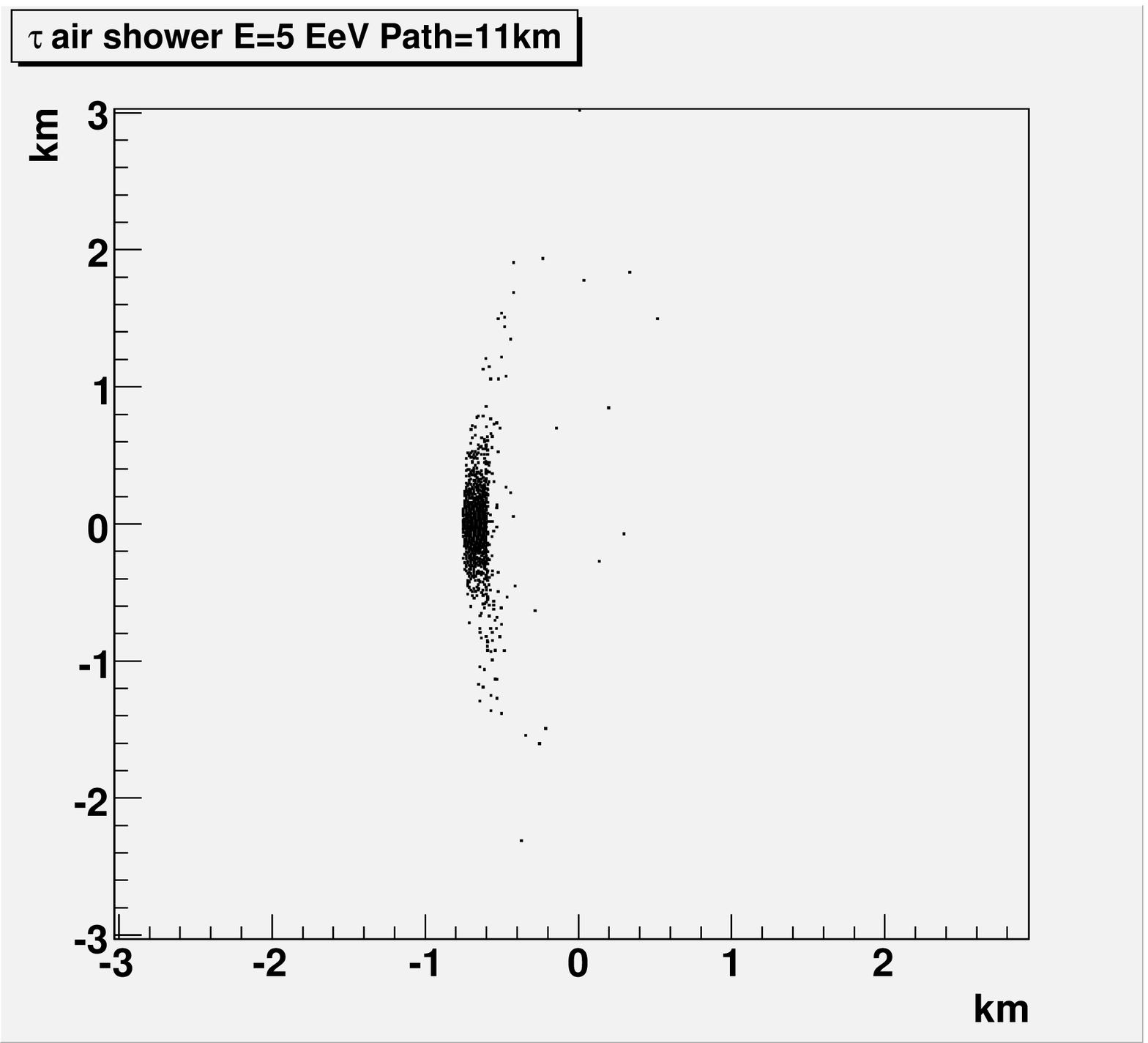}
\caption{
\label{fig:front}
Muon density produced in a time window of 500~ns at the detector location is estimated using the Aires simulation:
atmospheric (left) and $\tau$ showers (right) have
an energy of $5\times 10^{18}$eV and are coming from the right side of the plots.}
\end{center}
\end{figure}

The arrival time of the particles can provide information on the arrival direction of the shower.
The ordinary horizontal
hadron interaction has a plane wave front since it originates very far from the detector
while the $\tau$ shower is characterized by a
 front with a curvature radius $R\approx$10~km on detector plane
 (Fig.~\ref{fig:wavefront_sketch}).
Therefore, a time measurement of the shower development provides the
ability to identify
the direction of the reconstructed shower and the decay point.
The muon density generated at the detector array from atmospheric hadron showers
and from tau showers are compared using the Aires simulation (Fig.~\ref{fig:front}).
%(Figg.~\ref{fig:protshow},\ref{fig:taushow},\ref{fig:front}).
In the simulation, the hadron shower arrives at a zenith angle of 89$^\circ$,
while the $\tau$ shower emerges from the ground at a zenith angle $\theta=95^\circ$.
Both showers have an energy of $5\times 10^{18}$eV and are sampled in a time window of 500~ns.

Different strategies can be combined to distinguish
a $\tau$ shower from other particles originating from ordinary UHECRs:
\begin{itemize}
\item detector array pointing at different zenith angles (i.e. $\theta=92^\circ-95^\circ$) and leaving the possibility to change the azimuth angle
of each unit to enlarge the acceptance, \cite{crown} ;
\item measure the particle density which
is different for a ``new'' UHECR air shower
with a depth\footnote{The ``shower depth'' is the distance between the tau decay and the detector.}
of about 400~g/cm$^2$ ($R\approx$ 10~km),
and an ``old'' one with a depth larger than 4000~g/cm$^2$ ($R>$100~km);
\item measure the direction of the shower axis and its vertex;
\item measure the ratio between the short lived electromagnetic component and the long-lived muons of an air shower
(the possibility to distinguish between these two components is still under study).
\end{itemize}

\section{Tau shower reconstruction}
\label{sec:taurec}

A study of the tau shower reconstruction has been performed using Aires~\cite{Aires} and Corsika~\cite{Corsika} Monte Carlo simulations.
Tau shower Monte Carlo events with a zenith angle $\theta=95^\circ$ have been generated
and the detector array is assumed to lie on an inclined plane (i.e. mountain) of 45 degrees
and positioned 10~km away from the ground where the $\tau$ is emerging
. The conclusions are not strongly dependent on theta angle within 5 degrees.
In the simulation we consider muons and electrons with energy greater than
300 MeV and 20 MeV respectively.
The identification of the shower core, if it is inside the array, is estimated by calculating
the center of gravity of the shower with an accuracy of about 5~m.
This is affected by an additional uncertainty of approximately 50~m
due to the direction and inclination of the incoming shower, and to the exponential decrease of particle density.
With an angular resolution of less than $1^\circ$,
the intersection of the shower axis with the Earth's crust can be measured with an uncertainty of approximately 100~m.

The direction of the shower has been evaluated using a least squares fitting technique
for a function describing the position ($x,y,z$) of the front of the shower and the recorded arrival time $t_i$ at the detector site,
with respect to the time of decay ($t_0$) of the tau, by minimizing
$$ \chi ^{2} = \Sigma ^{N}_{i=1}(x_{i}l + y_{i}m + z_{i}k -c(t_{i}-t_{0}))^{2}$$
where the sum includes all the towers and $l$, $m$, $k$ are
the direction cosines.
Using the muons in the shower, the resulting zenith resolution is estimated to be of the order of $0.5^\circ$.
This result is independent on the size of the subsection of the detector array,
when an array grid of at least $100\times100$~m$^2$ with spacing $D=$25~m is set.
It should also be remarked that each tower in the detector array is ``blind'' to particles coming from a direction that
deviates of more than $7.5^\circ$ from the pointing axis.
Furthermore, if the shower is contained in the detector array, the shower front can be used to estimate
its radius, i.e. distance of the decay point of the tau (or shower depth),
and thus gain a further rejection of background events.

The possibility of detecting showers with the core outside the array will be
considered due to the large acceptance of units,
although the contamination due to background events may be larger.
If the core is outside of the array the reconstruction efficiency is
estimated to be of the same order even if not accurately evaluated.

\section{Trigger}
\label{sec:trigger}

The detector array is composed by several units spread over an area of 1km$\times$1km.
The remote location and the lack of easy communication links requires reliable low power electronics powered by solar panels,
and trigger logic at each tower.
Trigger algorithms, operating within each tower and between different towers,
must be developed to suppress lower energy cosmic ray showers and retain the events of interest.
A hierarchical event trigger can be used in order to keep the event rate within the limits of the data acquisition system.
Correlation of more towers can provide a further rejection of air showers.
Furthermore, the particle multiplicity from UHECRs is larger than in atmospheric showers.
The timing information of the front of the shower can also be used to discriminate between the events of interest with short radius ($R\approx10$km),
and the atmospheric air showers with much larger radius ($R>100$km).
Following these criteria it is possible to address different physics issues and select $\tau$ showers, muon bundles or
large zenith angle atmospheric interactions.

\section{Test results with a detector prototype at Jungfraujoch Station, Switzerland}
\label{sec:prototype}
Measurements were performed at the High Altitude Jungfraujoch Station (3600~m a.s.l.)
to understand detector characteristics and performance.
These studies are also aimed at the understanding of the possible sources of background
from inclined atmospheric showers at large zenith angles.
The environment of this test was intentionally chosen to be worse than at the final experimental proposed site.
At an altitude of 3600~m a.s.l. the electromagnetic component of vertical air showers is larger than at sea level,
increasing the probability of contamination from background events.
Studies of the detector performance with one module called
$tower$  at sea level were reported earlier~\cite{mine}.

\begin{figure}[htbp]
\begin{center}
\includegraphics[width=10cm,totalheight=6cm]{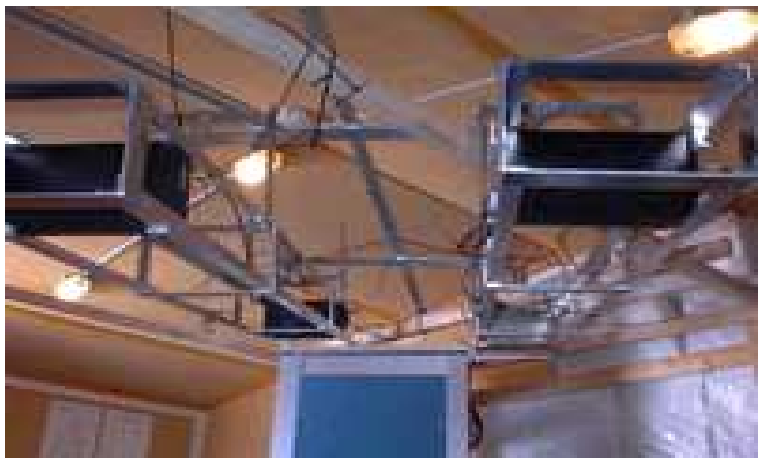}
\caption{\label{fig:detectorsetup}Two towers located at the Jungfrajoch High Altitude Station.}
\end{center}
\end{figure}

Towers with different tile sizes have been installed:
two  are instrumented with tiles of dimensions $12.5\times12.5\times 2$~cm$^3$
and are placed parallel to each other about 50~cm apart,
while another tower has tiles of $20\times 20\times 1.4$~cm$^3$
and was installed at a distance of 20~m from the other two (Fig.\ref{fig:detectorsetup}).
The use of tiles with different sizes is aimed at the optimization of detector acceptance and time resolution.
NIM and Camac modules with low threshold discriminators,
high resolution Time-to-Digital converter (TDC), and Analog-to-Digital converter (ADC)
were used to trigger the tracks and measure their time of flight.
The readout electronics used in these measurements
will be replaced by a customized board installed on each tower
for the final design of the detector array; using TDC-GPX by Acam, with its 80ps resolution,
and MATACQ by DAPNIA/CEA, capable of a continuous sampling at 1GHz for 4/8 channels, for a memory
depth of about 2500 points, it will be possible to implement a fully multi-hit solution for
time, charge and pulse shape analysis within a $2.5\mu s$ window.

\begin{figure}[htbp]
\begin{center}
\includegraphics[width=\textwidth]{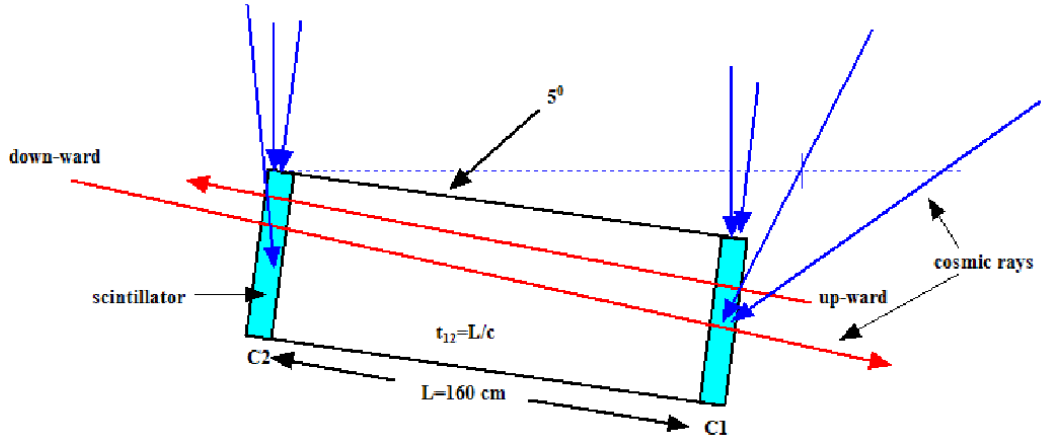}
\caption{\label{fig:tower} Schematic view of a tower (not to scale).
The up-going, down-going, and vertical tracks shown
can be distinguished using the time-of-flight information.}
\end{center}
\end{figure}

Each tower is composed of two parallel tiles (C1 and C2) of Kuraray
scintillator mounted on an aluminum frame (Fig.\ref{fig:tower}),
and are placed at a distance $L$ which can be varied from 1.4~cm to 200~cm.
The distance is set at $L$=160~cm
to distinguish between up-going and down-going tracks
and to reduce the background generated by ``mini-showers'',
due to particles traversing the scintillator near the edges~\cite{mine}.
The direction of the particles traveling through the tower can be determined
by the measurement of the time of flight between the two tiles.
The scintillators are read by a Hamamatsu H5873 photomultiplier tube (PMT) directly coupled
to the scintillator through an optical connector,
and without a light guide in order to minimize the time dispersion of the signal.
The PMT has a low-voltage power supply and a transit time spread of less than 0.8~ns.

\subsection{Detector performance}

\begin{figure}[htbp]
\begin{center}
\includegraphics[height=8cm,width=10cm]{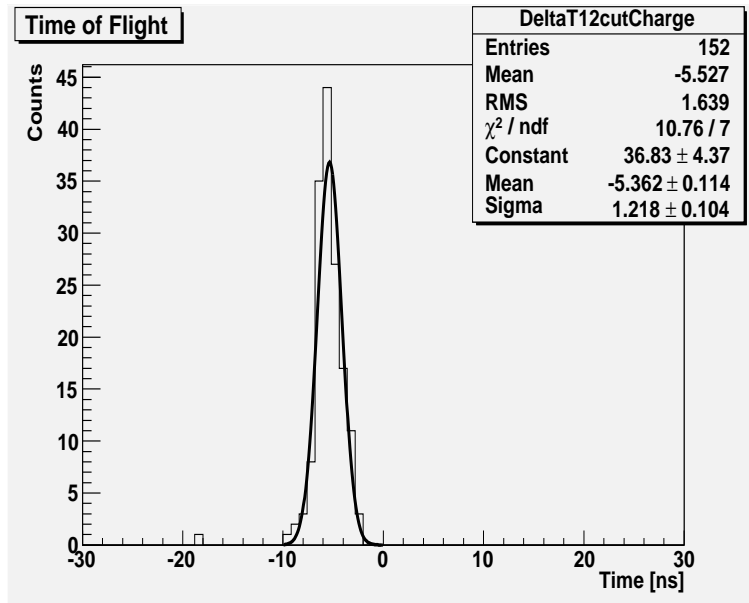}
\caption{\label{fig:tres} Difference of time-of-flight ($\Delta t_{12}$) between C1 and C2, when the $20\times20$~cm$^2$ tiles are 160~cm apart.
}
\end{center}
\end{figure}

A measurement aimed at determining the time resolution was performed using a single tower pointing
at a zenith angle $\theta=0^\circ$ and with a distance between tiles $L=$160~cm;
the difference of the signal arrival time at each of the two tiles, $\Delta t_{12}$,
was recorded (Fig.~\ref{fig:tres}).
The intrinsic time resolution of the tower is measured by the spread of the distribution $\sigma\simeq$1.2~ns.
A similar time resolution was obtained when the two tiles were placed at a distance $L=$1.4~cm.

In order to isolate the particles arriving longitudinally along the axis of the tower,
selection cuts are applied to the charge collected by the
two photomultipliers.
Orthogonally-traversing tracks deposit in the scintillator tiles approximately the same amount of energy,
within Landau fluctuations. Asymmetrically deposited charges characterize background tracks:
very low or very large amounts of deposited charge are associated to tracks traversing
near the edges of the scintillator or to multiple tracks, respectively.
These cuts were defined using the charge of vertical equivalent muons (CVEM).
Data were collected using a coincidence of the signals in C1 and C2 with a gate of 70~ns.
The light emitted by the scintillator has a cylindrical symmetry and the time resolution
is determined by the first photons that reach the PMT, without any reflection on the edges.
Photons hitting the corners of the tile will arrive at the PMT at a later time
and, due to the reflections, will have a smaller charge at the PMT's output.

%\begin{figure}[htbp]
%\begin{center}
%\includegraphics[width=12.cm,totalheight=10.cm]{cvem.eps}
%\caption{\label{fig:cvem}
%Charge in ADC counts deposited by a sample of muons crossing the two tiles at $0^\circ$ zenith angle.}
%\end{center}
%\end{figure}

%\begin{figure}[htbp]
%\begin{center}
%\includegraphics[width=12.cm,totalheight=10.cm]{figa2.eps}
%\caption{\label{fig:select}
%Charge in ADC counts deposited in the two scintillator tiles, C1 (top plot) and C2 (bottom plot),
%collected with a C1$\cdot$C2 coincidence within a time window of 3~ns around
%the signal region (blue) or outside of this time window (yellow).}
%\end{center}
%\end{figure}

\subsection{Measurements}

\begin{figure}[htbp]
\begin{center}
\includegraphics[width=12.cm,totalheight=10.cm]{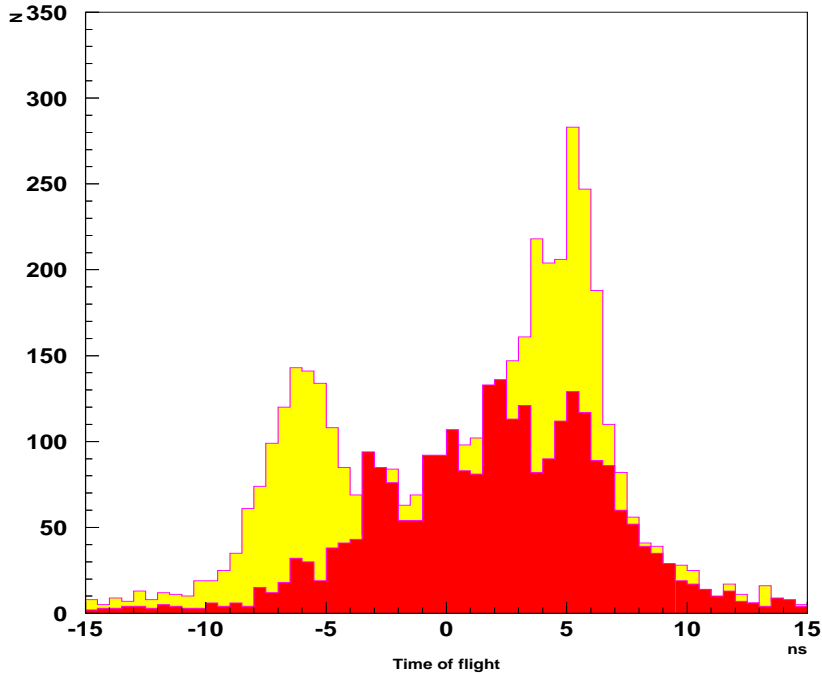}
\caption{\label{fig:pbnopb} Time-of-flight difference between two tiles $\Delta t _{12}$ in a tower
pointing at a zenith angle $95^\circ$ when a lead block of 3-cm thickness is placed in front of one tile (yellow shade) and no lead is present (red shade).
The two peaks at -5~ns and +5~ns correspond to up-going and down-going tracks, respectively.
The peaks at -3, 0 and 3 ns are due to parallel tracks, most likely vertical,
where one of them hits an ``inefficient region'', i.e. a corner, of one of the tiles.
}
\end{center}
\end{figure}

A measurement of the horizontal cosmic ray flux was performed using one tower with the tiles placed at a distance L=160~cm,
and pointing at a zenith angle $95^\circ$.
Thanks to its good time resolution, each tower is capable of distinguishing the particle direction between up-going and down-going tracks
by measuring the time-of-flight between the two tiles.
The two peaks at -5~ns and +5~ns
correspond to up-going and down-going tracks, respectively (Fig.~\ref{fig:pbnopb}).
The contamination of the events outside the two peaks is due to parallel tracks, most likely vertical,
or to tracks where the particle hits an ``inefficient region'', i.e. a corner, of one of the tiles.

The contamination in the flux above 90 degrees is due to muon decays and
can be eliminated by absorber material (lead) or, alternatively, by correlating information from distant towers.
The suppression provided by the lead block is different for up-going and down-going tracks,
and it is due to the shadow of the mountain which remove part of
the down-going track flux.

\begin{figure}[htbp]
\begin{center}
\includegraphics[width=12.cm,totalheight=10.cm]{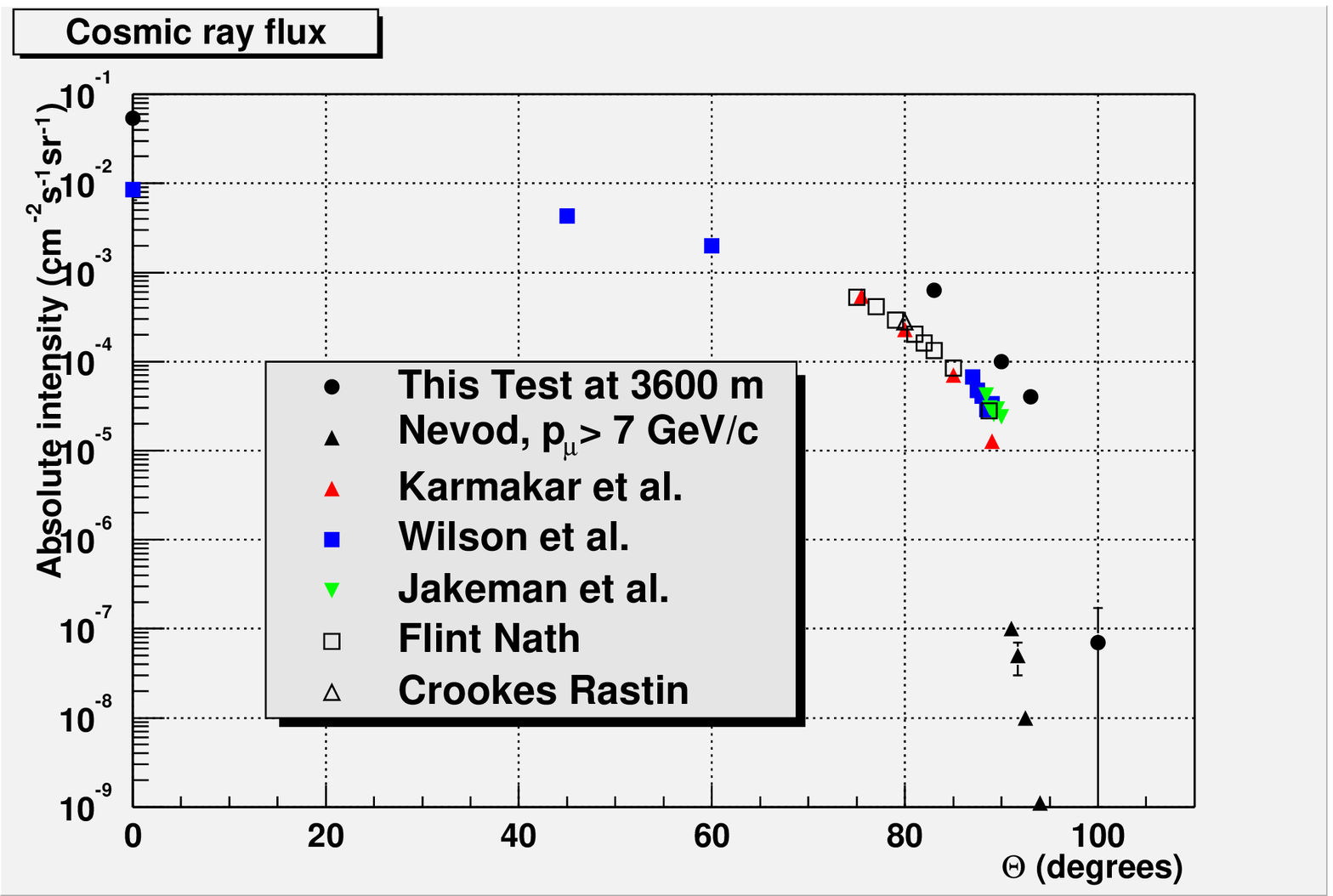}
\caption{\label{fig:fluxes} Measurements of fluxes at several zenith angles in the
High Altitude Jungfraujoch Station compared with results from other experiments at sea level,
\cite{Nevo,Karma,Wilson,Jake,Flint,Crookes}.}
\end{center}
\end{figure}

A measurement of
the cosmic ray flux was performed using two  towers pointing at different zenith angles
and compared to results from other experiments (Fig.~\ref{fig:fluxes}).
The detector sensitivity is $10^{-7}$cm$^{-2}$sec$^{-1}$sr$^{-1}$ at $\theta=98^\circ$.

\section{Conclusions}

The study and the implementation of a detector array designed to measure the flux from UHE neutrinos has been discussed.
Results using a detector prototype have also been presented.
In the hypothesis of full mixing of $\nu _{\mu} \to  \nu _{\tau}$,
after two years of data-taking with the full detector array,
about 10 events could be collected in the energy range
$10^{17}-10^{20}$eV  from AGN tau neutrinos,
or an upper limit on the cosmic ray flux, as well as measurements of muon bundles
at large zenith angles, could be achieved for other scenarios.
Assuming that the neutrino energy spectrum follows a Fermi-like power law
$E^{-2}$, the sensitivity with 3 years
of observation is estimated to be about 60 eV cm$^{-2}$s$^{-1}$sr$^{-1}$
in the energy range $10^{17-20}$ eV.
This value would provide competitive upper limit with present and future
experiments. We found also that, in the same time, this system can observe about one GZK neutrino
event per km$^2$.
Even if we have not investigated accurately cases in which the core of
the shower is outside of the array, we expect the reconstruction efficiency
not to decrease significantly.
The proposed detector array is complementary to other experiments in this
energy range, like Amanda or Antares. It is also more efficient than those experiments
using ``double-bang'' or ``lollipop'' strategies which suffer from the inefficiency due to the need
of detecting the hadronic shower from the initial $\nu _{\tau}$,
where the $\tau$ track and a second cascade must be close to the detector.
In terms of energy range, it is comparable to the Auger and NuTel experiments
but the former has no large detection efficiency for taus emerging
from the ground and the latter has only 10\% of the duty cycle.
A recent proposal that uses a fluorescence technique~\cite{Cao}
will be competitive with this project in terms of detected events if
two new telescopes pointing at a long chain of mountains are built.

\section*{Acknowledgments}
During the writing of this paper we have benefited from
discussions with several colleagues; among them J. Russ,
J.F. Beacom, M.A. Huang .

\end{document}